\documentclass[useAMS, usenatbib, usegraphicx]{mn2e}
\usepackage{times}

\title[Star Formation Around KR~140 ]{Sequential and Spontaneous Star
  Formation Around the Mid-Infrared Halo \mbox{H\,{\sc ii}} Region KR~140}

\author[Kerton et al.]{C. R. Kerton$^1$\thanks{E-mail:
    kerton@iastate.edu}, K. Arvidsson$^1$, Lewis B. G. Knee$^2$, and
  C. Brunt$^3$ \\
  $^1$Department of Physics and Astronomy, Iowa State University,
    Ames, IA 50011, USA\\
  $^2$Atacama Large Millimeter Array, Avenida El Golf 40, Piso 18,
    Las Condes, Santiago, Chile\\
  $^3$School of Physics, University of Exeter, Stocker Road, Exeter
    EX4 4QL, UK}

\begin{document}

\maketitle

\begin{abstract}
We use 2MASS and \emph{MSX} infrared observations, along with new
molecular line (CO) observations, to examine the distribution of young
stellar objects (YSOs) in the molecular cloud surrounding the halo
\mbox{H\,{\sc ii}} region KR~140 in order to determine if the ongoing
star-formation activity in this region is dominated by sequential star
formation within the photodissociation region (PDR) surrounding the
\mbox{H\,{\sc ii}} region. We find that KR~140 has an extensive population
of YSOs that have ``spontaneously'' formed due to processes not related to the
expansion of the \mbox{H\,{\sc ii}} region. Much of the YSO population
in the molecular cloud is concentrated along a dense filamentary
molecular structure, traced by C$^{18}$O, that has not been erased by
the formation of the exciting O star. Some of the previously observed
submillimetre clumps surrounding the \mbox{H\,{\sc ii}} region are
shown to be sites of recent intermediate and low-mass star formation
while other massive starless clumps clearly associated with the PDR
may be the next sites of sequential star formation.
\end{abstract}


\section{Introduction} \label{sec:intro}

This study is the fourth in a series of papers exploring the
structure of the \mbox{H\,{\sc ii}} region KR~140 and its associated 
star-forming activity. In \citet*{ker99} VES~735, the O-star
powering the \mbox{H\,{\sc ii}} region was examined. A second paper, 
\citet*{bal00}, was a multiwavelength study of the structure,
energetics, and kinematics of the \mbox{H\,{\sc ii}} region. Finally, 
\citet{ker01} presented an analysis of submillimetre (submm hereafter)
observations of the region at 450 and 850~$\mu$m. The main result of
\citet{ker01} was the discovery of numerous submm dust cores located
within the molecular gas surrounding the \mbox{H\,{\sc ii}} region,
including a large number of cores that were clearly located at the
interface between ionized and molecular gas, a likely location for
star-formation induced or ``triggered'' by the expansion of the
\mbox{H\,{\sc ii}} region \citep{elm98}. Two of the more isolated
cores were clearly associated with \emph{IRAS} sources, but the low
resolution of \emph{IRAS} combined with the extensive diffuse emission of
dust associated with the \mbox{H\,{\sc ii}} region made it impossible
to determine if any of the other cores were associated with
star-formation activity. 

The goal of this study is to determine the distribution of young
stellar objects (YSOs) throughout the KR~140 molecular cloud. We focus
especially on the embedded stellar content of the submm cores as these
are potentially the youngest star-forming regions. We use the
spatial distribution of the YSOs relative to the photodissociation
region (PDR) surrounding the \mbox{H\,{\sc ii}} region as a means of
gauging the relative importance of spontaneous and sequential, or
triggered, star formation in this region. To achieve this we have
analyzed a combination of submm data from \citet{ker01} and
\citet{moo07}, 2MASS (2 Micron All Sky Survey) and Midcourse Space
Experiment (\emph{MSX}) infrared data, and newly acquired $^{12}$CO,
$^{13}$CO and C$^{18}$O molecular line data from the Five College
Radio Astronomy Observatory (FCRAO). In the next section we review the
pertinent parameters of KR~140 and provide information on the new
observations of the region. In \S~\ref{sec:2mass} and
\S~\ref{sec:dist} we first show how 2MASS data can be utilized to
identify the YSO population and then examine its spatial
distribution. Our results are discussed in \S~\ref{sec:discuss} and
conclusions are presented in \S~\ref{sec:conclude}.

\section{KR~140: Properties and Observations} \label{sec:kr140}
KR~140 \citep[$l=133\fdg425$, $b=+0\fdg054$;][]{kal80} is a 5.7~pc
diameter \mbox{H\,{\sc ii}} region located at a distance of
$2.3\pm0.3$~kpc. It is close to, but apparently isolated from, the
large W3/W4/W5 star-formation complex. The region is ionized by a
single O8.5 V(e) star VES~735. Foreground extinction is $A_V =
5.7\pm0.2$ magnitudes toward VES~735 and ranges between $A_V \sim 6 -
7$ over the \mbox{H\,{\sc ii}} region. Molecular line observations
($^{12}$CO J$=$1$-$0) indicate that the \mbox{H\,{\sc ii}}
region is associated with a molecular cloud with a mass in the range
10$^{3.7}$ -- 10$^{4.0}$ M$_\odot$. A detailed discussion and derivation
of these properties can be found in \citet{ker99} and \citet{bal00}.

\subsection{Submillimetre Observations}\label{sec:submm_obs}

For this study we used a combined sample of submm clumps identified in
the \citet{ker01} and \citet{moo07} 850~$\mu$m SCUBA scan-maps of the
KR~140 region. There are only minor disagreements between the two samples
where they overlap due to the different ways in which
structures were identified in the two studies and the correspondence
between the two samples is shown in Table~\ref{tbl:subcomp}. In the
area covered by our molecular-line observations
($2^{\rmn{h}}18^{\rmn{m}} 15^{\rmn{s}} \leq \alpha_{\rmn{2000}} \leq
2^{\rmn{h}} 22^{\rmn{m}} 30^{\rmn{s}}$  and $60\degr 54\arcmin \leq
\delta_{\rmn{2000}} \leq 61\degr 20\arcmin$, see
\S~\ref{sec:co}) there are a total of 39 submm clumps: KMJB 1 - KMJB
20 from \citet{ker01} and clumps 7, 9, 12-13, 21, 39-40, 42-45 and
47-54 from \citet{moo07}. 

\begin{table}
\caption{Spatial correspondence between \citet{ker01} and \citet{moo07} submm clumps (KMJB and M07 respectively).}
\label{tbl:subcomp}
\centering
\begin{tabular}{ccccc}
\hline
KMJB & M07 &  & KMJB & M07  \\
\hline
1 &  37       &  & 11 & 14       \\   
2 &  $\cdots$ &  & 12 & 11       \\
3 &  32       &  & 13 & 8 \& 10  \\
4 &  33       &  & 14 & 5        \\
5 &  26 \& 29 &  & 15 & $\cdots$ \\
6 &  25       &  & 16 & 6        \\
7 &  24       &  & 17 & $\cdots$ \\
8 &  23       &  & 18 & 4        \\
9 &  17       &  & 19 & 3        \\
10 & $\cdots$ &  & 20 & $\cdots$ \\
\hline
\end{tabular} 
\end{table}

\subsection{Mid-Infrared \emph{MSX} Observations}\label{sec:msx}

Mid-infrared emission from polycyclic aromatic hydrocarbons (PAHs) can
be used to detect the PDRs that arise at the
interface between ionized gas in an \mbox{H\,{\sc ii}} region and surrounding
molecular material \citep{gia94}. We obtained \emph{MSX} Band A ($\lambda_o =
8.3~\mu$m, $\lambda=7-11~\mu$m) images of KR~140 with $20\arcsec$
resolution taken as part of the \emph{MSX} Galactic Plane Survey
\citep{pri01}. The Band A emission is dominated by intense line
emission from PAHs caused by various C$-$C stretching and C$-$H
bending modes. An up-to-date listing of the numerous PAH features in
the mid-infrared can be found in \citet{dra07}.

\begin{figure}
\includegraphics[width=84mm]{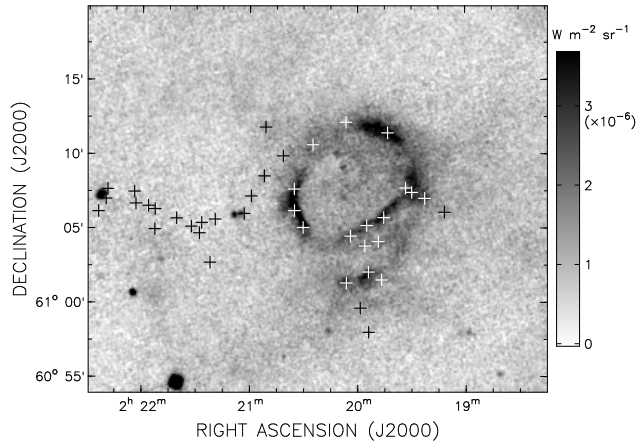}
\caption{\emph{MSX} Band A (8.3~$\mu$m) image of the KR~140 region. Note the
  distinctive halo morphology of the PDR. White crosses indicate the
  positions of submm clumps apparently associated with
  the PDR and black crosses indicate the submm clumps found away from the
  PDR.}
\label{fig:msx}
\end{figure}

The distinctive mid-infrared morphology exhibited by KR~140 (see
Fig.~\ref{fig:msx}) and in other similar \mbox{H\,{\sc ii}}
regions \citep[e.g.][]{coh01,deh03} arises from changes in the
relative balance between ultraviolet (UV) photons which
preferentially destroy PAHs and those which excite their
emission. Deep within the \mbox{H\,{\sc ii}}  region, hard UV
radiation destroys PAHs and are themselves destroyed. Outside the
PAH-free zone, the remaining UV photons move into the surrounding PDR
and excite PAH emission in a layer around the \mbox{H\,{\sc ii}}
region. This layer is relatively thin because UV fluxes continue to
drop through the PDR, resulting in a drop in PAH emission. The
overall result is a striking ring-like or ``halo'' morphology. In
Fig.~\ref{fig:msx} we also show the location of the submm clumps
discussed in \S~\ref{sec:submm_obs}: the correspondence between the location
of some of the cores and the interface region traced by the
mid-infrared emission is evident. 

\subsection{Molecular Line Observations}\label{sec:co}

\begin{figure}
\includegraphics[width=84mm]{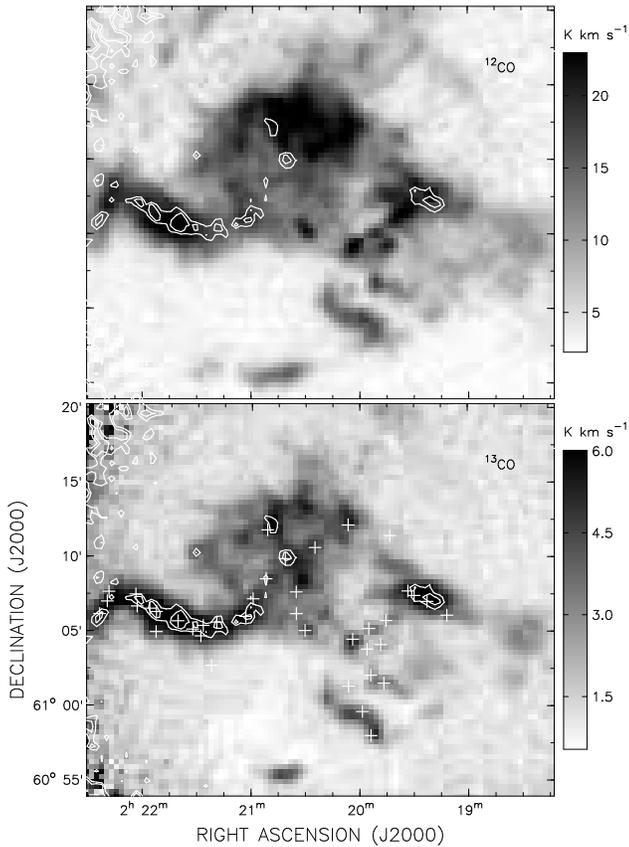}
\caption{Integrated $^{12}$CO (top) and $^{13}$CO (bottom) maps of the KR~140
  molecular cloud. Contours in both panels show integrated C$^{18}$O emission
  at the 1.2 and 1.4~K~km~s$^{-1}$ levels. White crosses in the lower
  panel indicate the positions of all detected submm clumps. 
\label{fig:comaps}}
\end{figure}

We obtained new molecular line observations of KR~140 in order to
determine the full spatial extent of the surrounding
molecular cloud and to identify any high column density regions.
 Maps, covering a $0\fdg8$ square region around KR~140, were
obtained in the J$=$1$-$0 transition of $^{12}$CO (115.3 GHz),
$^{13}$CO (110.2 GHz) and C$^{18}$O (109.8 GHz) using the SEQUOIA
array on the 14~m FCRAO millimetre-wave telescope \citep{eri99}. The
resulting data cubes have a spatial resolution of $\sim 45\arcsec$
($20\arcsec$ pixel grid) and a velocity resolution of
$0.065$~km~s$^{-1}$. Integrated intensity (zeroth moment) $^{12}$CO
and $^{13}$CO maps of emission between $-54 <$ V$_\mathrm{LSR}$ $<
-44$~km~s$^{-1}$ are shown in Fig.~\ref{fig:comaps}. Contours of
integrated intensity C$^{18}$O emission are also shown on each map
highlighting the two main regions of high column density found to the
east and west of the \mbox{H\,{\sc ii}} region. 

We find a mass of $8500\pm1000$ M$_\odot$ for the cloud based on the
integrated $^{12}$CO data. A value of $X = 1.9\times10^{20}$ from
\citet{str96} to convert between integrated intensity (K~km~s$^{-1}$)
and H$_2$ column density and using a mean  molecular weight of 2.29 to
account for the molecular nature of the  hydrogen and the presence of
He. This estimate is larger than that quoted in \citet{bal00} mainly because we
are using a larger spatial range to define the associated cloud.

\subsection{Near-infrared Data}\label{sec:nir}

Photometric (\emph{J},\emph{H},\emph{K$_{\rm{s}}$}\footnote{In this
study we use \emph{K$_{\rm{s}}$} and \emph{K} band photometry
interchangeably}) data on YSOs associated with KR~140 were obtained
from the 2MASS  All-Sky Point Source Catalog (2MASS PSC hereafter)
using the on-line {\sc gator} query tool at the Infrared Science Archive
(IRSA).  Detailed discussion of the survey and
the various data products are available in \citet{skr06} and
\citet{cut03}. Here we note that the overall quality of each 2MASS
photometric measurement is indicated by a photometric quality flag
(``ph\_qual'') in the 2MASS PSC. Valid measurements are indicated by a
ph\_qual value of A, B, C, or D for each filter with the
signal-to-noise of the detection decreasing as one moves from ph\_qual
= A through ph\_qual = D. Non-detection in a filter is typically
indicated by ph\_qual = U, in which case an upper brightness limit is given. 

\section{Using 2MASS to Define the YSO Population} \label{sec:2mass}

\subsection{\emph{JHK} Colour-Colour and Colour-Magnitude Diagrams}
\label{sec:cccm}

Since we know the distance to KR~140 and the amount of foreground extinction,
our approach is to scale observations of different types of YSOs in
local star-forming regions to the distance of KR~140, including the
effect of foreground extinction, thus defining regions on the
\emph{JHK} colour-colour (CC) and colour-magnitude (CM) diagrams where
we would expect to find YSOs associated with the KR~140 molecular cloud. 

\begin{figure}
\includegraphics[width=84mm]{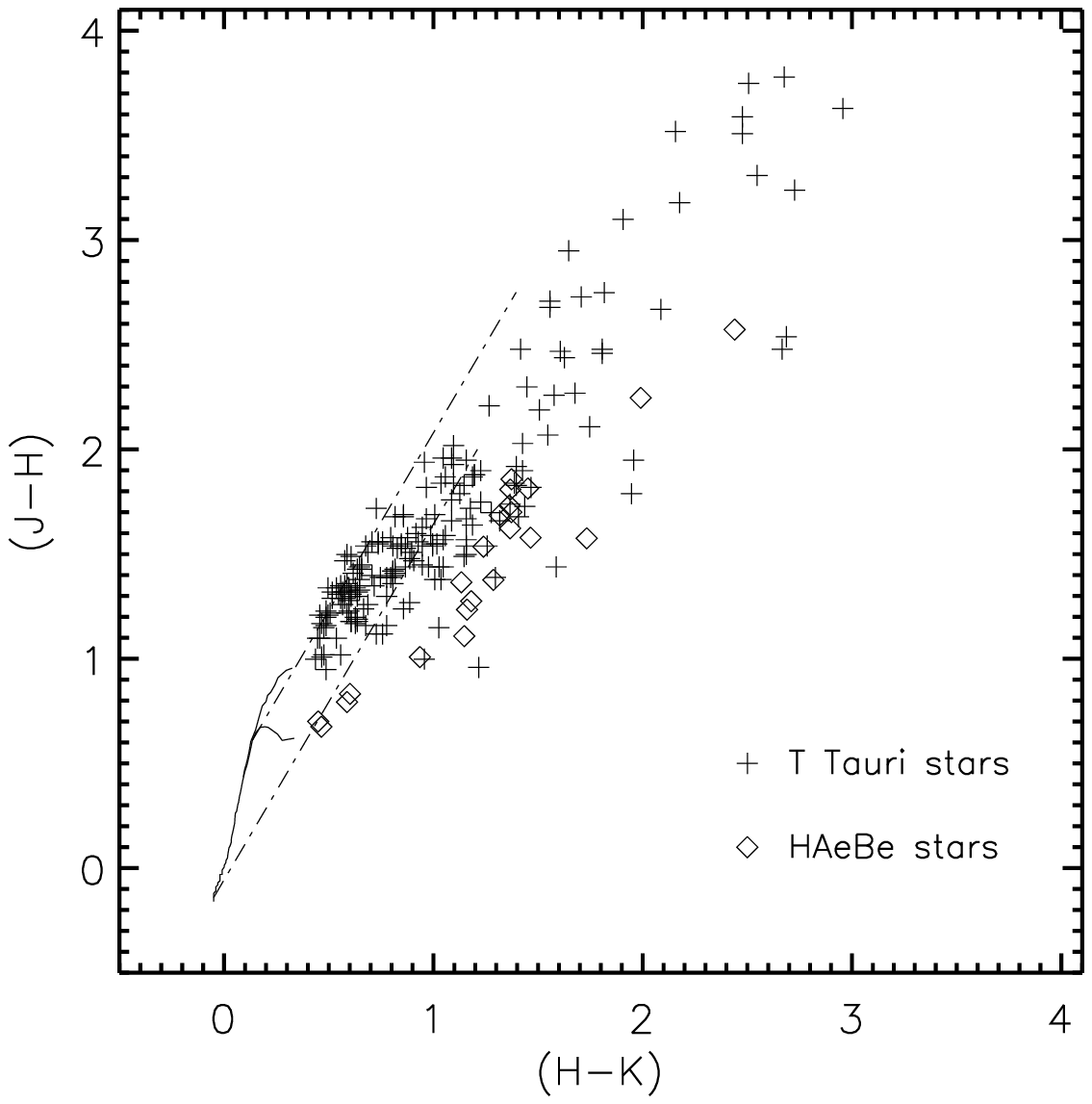}
\caption{Near-infrared colour-colour diagram. T Tauri and HAeBe data
  from \citet{ken95} and \citet{the94} respectively are shown shifted
  to a distance of 2300 pc and with $A_V = 5.5$ of foreground
  reddening. The solid lines show the intrinsic colours of
  main-sequence (V) and giant (III) stars \citep{koo83,bes88}. The
  dash-dot lines show $A_V = 20$ reddening vectors for a K5~V and O9~V
  star derived using the infrared extinction law
  \emph{E(J$-$H)/E(H$-$K)} $= 1.7\pm0.4$ from \citet{rie85}. \label{fig:ccd}}
\end{figure}

In Fig.~\ref{fig:ccd} we have plotted near-infrared colours of
T~Tauri stars (approximately solar-mass YSOs with optically thick
disks of circumstellar material) from the \citet{ken95} study of the
Taurus-Auriga region (distance $\sim 140$~pc). Intermediate-mass YSO
colours are illustrated using 2MASS photometry for a subset of the
\citet*{the94} catalogue of Herbig Ae and Be stars (HAeBe) that have distance
estimates from \citet{fin84}. One sees that the majority of the sample
lies outside of the band of reddened stellar photospheres, with a
portion of the T~Tauri sample being the main exception. The region
occupied by our YSO sample agrees nicely with the CC diagrams
constructed using an extensive grid of YSO models presented in
\citet{rob06}, once the appropriate amount of foreground extinction is
applied to the model-based CC diagrams.

\begin{figure}
\includegraphics[width=84mm]{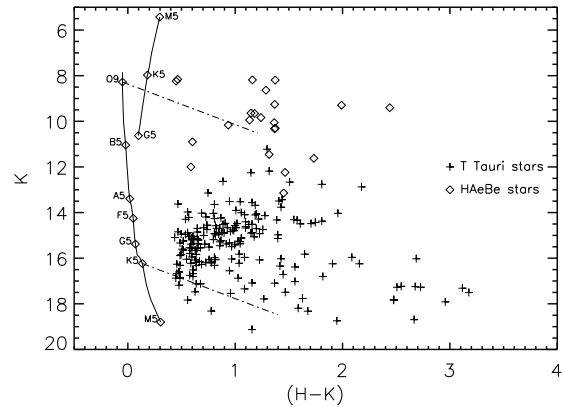}
\caption{Colour-magnitude diagram for objects at a distance of
  2300~pc. Data are the same as in Fig.~\ref{fig:ccd}.
  The main-sequence (V) and giant branch (III) are shown (solid lines)
  with some representative spectral types labeled. Reddening vectors
  (dash-dot lines) are drawn for $A_V = 20$ for an O9~V and K5~V star.
\label{fig:cmd}}
\end{figure}

The near-infrared CM diagram showing scaled T Tauri and HAeBe data is
shown in Fig.~\ref{fig:cmd}. The utility
of having both the CC and CM diagrams available for data analysis is
readily apparent. For example, in the CM diagram the HAeBe sample is
clearly distinct from the lower luminosity T~Tauri stars. Also, possible
confusion between HAeBe stars and giants can be avoided by using the
CC diagram, where the two samples are well separated.

\begin{figure}
\includegraphics[width=84mm]{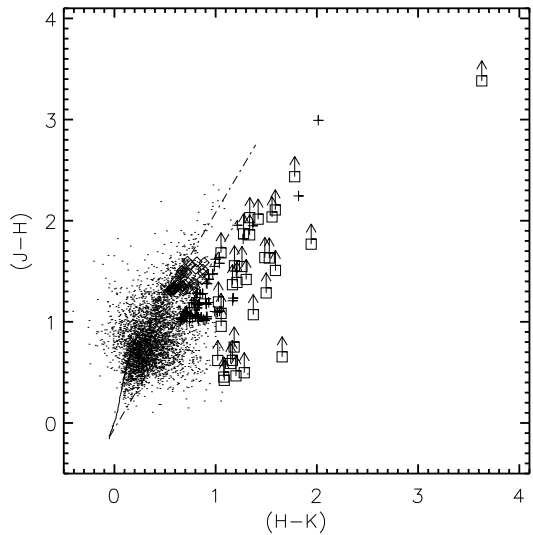}
\caption{2MASS Colour-colour diagram. Colours for the 4144 stars with
  valid \emph{JHK} photometry are plotted (dots) along with colours of
  the P1 (crosses), P1+ (diamonds), and P2 (squares) YSO
  samples. The P2 (\emph{J}$-$\emph{H}) colours are lower limits
  (indicated by the arrows). Note, for clarity, error bars are
  \emph{not} shown. 
\label{fig:ccd_p123}}
\end{figure}

\begin{figure}
\includegraphics[width=84mm]{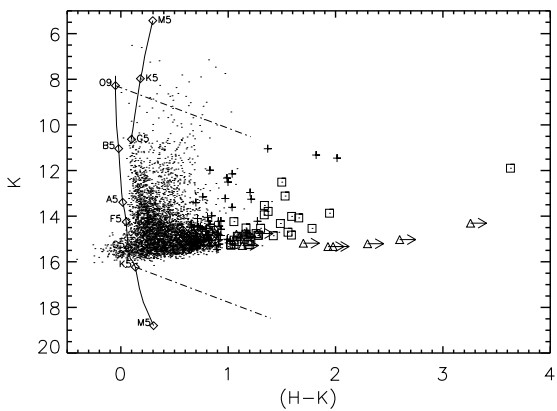}
\caption{2MASS Colour-magnitude diagram. As Fig.~\ref{fig:ccd_p123}
  but also showing the P3 (diamond) YSO sample. The P3
  (\emph{H}$-$\emph{K}) colours are lower limits (indicated by the arrows).
\label{fig:cmd_p123}}
\end{figure}

Table~\ref{tbl:ysomag} shows the average magnitude and standard
deviation along with the maximum and minimum magnitude for each of the
YSO samples. Also tabulated are the 2MASS limiting magnitudes for a
complete sample (``10$\sigma$'') and for ph\_qual = D quality
photometry (``D''; see \S~\ref{sec:nir}). We see that HAeBe stars will
be detected in all three bands within the 10$\sigma$ limits. T~Tauri
stars will also be detected in all three bands although the sample
will be incomplete, missing very heavily embedded sources

\begin{table}
\caption{\emph{JHK} magnitudes of YSO samples at 2300~pc.}
\label{tbl:ysomag}
\begin{tabular}{lccccccc}
\hline
Sample  & Band & Avg. & $\sigma$ & min & max & 10$\sigma^{a}$ & D$^{b}$ \\
\hline
HAeBe   & \emph{J} & 12.73 & 1.8 &  9.30 & 16.41 & 15.8 & 17.2 \\
        & \emph{H} & 11.28 & 1.5 &  8.62 & 14.59 & 15.1 & 16.1 \\
        & \emph{K} & 10.03 & 1.4 &  8.16 & 13.14 & 14.3 & 15.3 \\
T Tauri & \emph{J} & 18.21 & 2.0 & 13.91 & 25.73 & 15.8 & 17.2 \\
        & \emph{H} & 16.50 & 1.6 & 12.52 & 21.36 & 15.1 & 16.1 \\
        & \emph{K} & 15.44 & 1.3 & 11.22 & 19.12 & 14.3 & 15.3 \\
\hline
\end{tabular}
$^{a}$2MASS completeness limits.\\
$^{b}$2MASS ph\_qual = D limits for KR~140 region.\\
\end{table}

\subsection{The YSO Population of KR~140}\label{sec:ysopop}

\begin{figure}
\includegraphics[width=84mm]{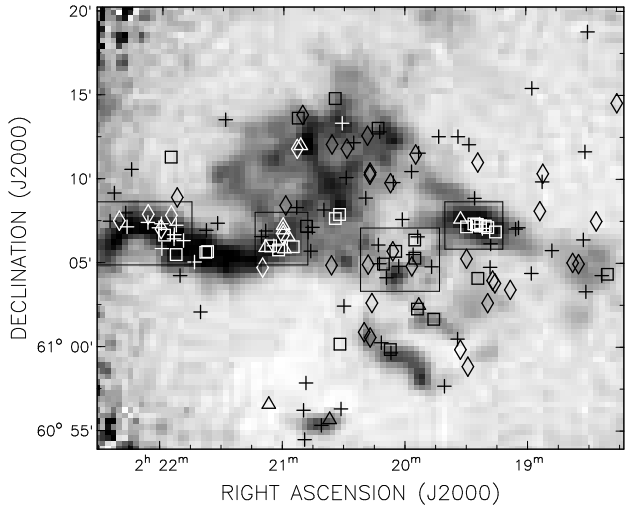}
\caption{The spatial distribution of the P1 (crosses) P1+
  (diamonds), P2 (squares) and P3 (triangles) YSO samples are shown
  against the integrated $^{13}$CO emission associated with
  KR~140. The boxes show the location of the four concentrations of
  YSOs discussed in the text, Region A-D, from left to right.  White or
  black symbols are used interchangeably depending on the background
  level.
\label{fig:p123dist}}
\end{figure}

Using the integrated CO maps as a guide we investigated a
$0\fdg5\times0\fdg4$ region centered on the KR~140 \mbox{H\,{\sc ii}} region
($2^{\rmn{h}} 18^{\rmn{m}} 15^{\rmn{s}} \leq \alpha_{\rmn{2000}} \leq
2^{\rmn{h}} 22^{\rmn{m}} 30^{\rmn{s}}$  and $60\degr 54\arcmin \leq
\delta_{\rmn{2000}} \leq 61\degr 20\arcmin$). The 2MASS PSC was
queried using the {\sc gator} software at IRSA. All of the
catalogue searches and the criteria used to identify YSOs discussed
below are summarized in Table~\ref{tbl:cats}. Figs.~\ref{fig:ccd_p123}
and \ref{fig:cmd_p123} show both the YSO and non-YSO population of the
region for comparison while Fig.~\ref{fig:p123dist}
shows the spatial location of the YSO sample. 

We first searched for all stars with valid photometry (i.e., ph\_qual
values of A, B, C, or D in all three bands\footnote{ph\_qual MATCHES
  [A-D][A-D][A-D] in {\sc SQL} used by {\sc gator}}). This search returned 4144
stars from the catalogue. The surface density of stars is fairly uniform
across the search area at $\sim 4.7$ stars~/$\sq'$ ($1\sigma =
1.3$). Within this sample of stars we searched for YSOs defined as
stars with (\emph{J}$-$\emph{H}) $> 1$ and (\emph{J}$-$\emph{H}) $<
1.7$ (\emph{H}$-$\emph{K}) $-$ 0.075. The first limit on
(\emph{J}$-$\emph{H}) reflects the fact we expect there to be
sufficient foreground extinction to push any YSO above the
(\emph{J}$-$\emph{H}) $> 1$ line. The second criteria selects stars
lying redward of the reddening vector associated with an O6~V
star. This search finds 72 stars which we call the P1 sample. 

We did an additional search for stars that may be located in the overlap
region of T~Tauri stars and reddened stellar photospheres. This region
is located between $1.3 <$ (\emph{J}$-$\emph{H}) $< 1.6$ and between the
two main-sequence reddening vectors. We also required that K $> 14.5$ in
order to reduce contamination from giant stars and distant early
main-sequence stars. This results in an
additional 38 stars, which we call the P1+ sample. We note the
conclusions of this paper are not sensitive to whether or not one uses
the more conservative selection criteria of the P1 sample or the
combined P1 and P1+ sample of 110 stars. We expect that any
contamination of the larger sample will arise from distant A and F
main-sequence stars as closer, later spectral type stars will tend not
to have enough extinction to match the colour criteria.

The KR~140 region was then re-examined for stars with an upper
brightness limit at \emph{J} and valid \emph{H} and \emph{K}
photometry (ph\_qual = [U][A-D][A-D]). This catalogue search returned 133
stars. Possible YSOs were defined using \emph{(H$-$K)} $> 1$ and
(\emph{J}$-$\emph{H}) $< 1.7$ \emph{(H$-$K)} $-$ 0.075. The latter
constraint places the (\emph{J}$-$\emph{H}) lower limit below the
reddened stellar photosphere region of the CC diagram. In total 33
stars match these criteria and we identify them as the P2
sample. 

Finally we searched the region for stars with only upper brightness
limits in \emph{J} and \emph{H} along with valid \emph{K} band
photometry (ph\_qual = [U][U][A-D]). Of the 37 stars found, 10 of them
were defined as possible YSOs by requiring the lower limit of
\emph{(H$-$K)} $> 1$. 

\begin{table}
\caption{YSO samples.}
\label{tbl:cats}
\begin{tabular}{lccc}
\hline
Sample & Number & ph\_qual values & Colour limits$^{a}$ \\
\hline
P1  & 72 & [A-D][A-D][A-D] &  \emph{(J$-$H)} $>1$                  \\
    &    &                 &  \emph{(J$-$H)} $<1.7$\emph{(H$-$K)}$-0.075$\\
P1+ & 38 & [A-D][A-D][A-D] &  $1.3<$ \emph{(J$-$H)} $<1.6$ \\
    &    &                 &  \emph{(J$-$H)} $< 1.7$\emph{(H$-$K)}$-0.075$\\
    &    &                 &  \emph{(J$-$H)} $< 1.7$\emph{(H$-$K)}$+0.3805$\\
    &    &                 &  $K>14.5$ \\
P2  & 33 & [U][A-D][A-D]   &  \emph{(J$-$H)$_L$} $<1.7$\emph{(H$-$K)}$-0.075$\\
    &    &                 &  \emph{(H$-$K)} $>1.0$ \\
P3  & 10 & [U][U][A-D]     &  \emph{(H$-$K)$_L$} $> 1.0$  \\
\hline
\end{tabular}
$^{a}$\emph{(J$-$H)$_L$} and \emph{(H$-$K)$_L$} indicate lower limits. 
\end{table}

\section{The YSO Population of the KR~140 Molecular Cloud} \label{sec:dist}

Using the 2MASS sample defined above, we now explore two
questions: 1) What is the association between the YSOs and the submm
sources identified by \citet{ker01} and \citet{moo07}, and 2) What is
the overall spatial distribution of the YSOs through the molecular cloud?

\subsection{YSOs and Submm Sources} \label{sec:submm}

The submm clumps identified by \citet{ker01} have typical diameters of
0.5~pc (10$^5$~AU), which is about an order of magnitude larger than
the 0.05~pc (10$^4$~AU) size scale typically associated with the
smaller core structures within molecular clouds thought to be
forming individual stars, for example, the \citet*{kir06}
observations of the Perseus molecular cloud and the Class 0/I YSO models of
\citet{rob06}. Thus it is quite possible that a single clump will
have multiple sites of star-formation associated with it and that
these regions need not correspond with the peak position of the submm
emission. We examined our entire YSO sample for positional
associations. For the \citet{ker01} sample we used a search radius of
D$_\rmn{eff}$/2, where D$_\rmn{eff}$ is the deconvolved FWHM of a
Gaussian fit to an azimuthally-averaged profile of each source. The
average search radius was 0.18~pc. Since we do not have size
information for the \citet{moo07} sample, we used a 20$\arcsec$ search
radius (0.22~pc at 2.3~kpc) for each
source in that sample. Table~\ref{tbl:ysosubmm} shows the results of this
search. Column~1 gives the clump ID number. Columns~2 through 5
contain the 2MASS PSC source number for associated YSOs from the P1,
P1+, P2, and P3 samples respectively. Column~7 contains the column
density of the submm core expressed in terms of visual extinction,
$A_V$. For the \citet{moo07} samples masses were calculated for a
distance of 2.3~kpc and a clump temperature of 20~K assuming that all
the clumps had a diameter of 0.36~pc. This results in $A_V \sim 10
S_{850}$ where $S_{850}$ is the integrated flux density at
850~$\mu$m. All of these $A_V$ estimates are very rough as the column
density will vary with the square of the estimated diameter and the
temperature enters in through the exponential in the Planck
function. Values for the remaining cores were taken from
\citet{ker01}. Finally, Column~7 denotes if the core is apparently
isolated or associated with the PDR. The table includes two submm
clumps, 3 and 53m, that have only associated \emph{IRAS} sources and
are possible Class 0/I YSOs.

\begin{figure}
\includegraphics[width=84mm]{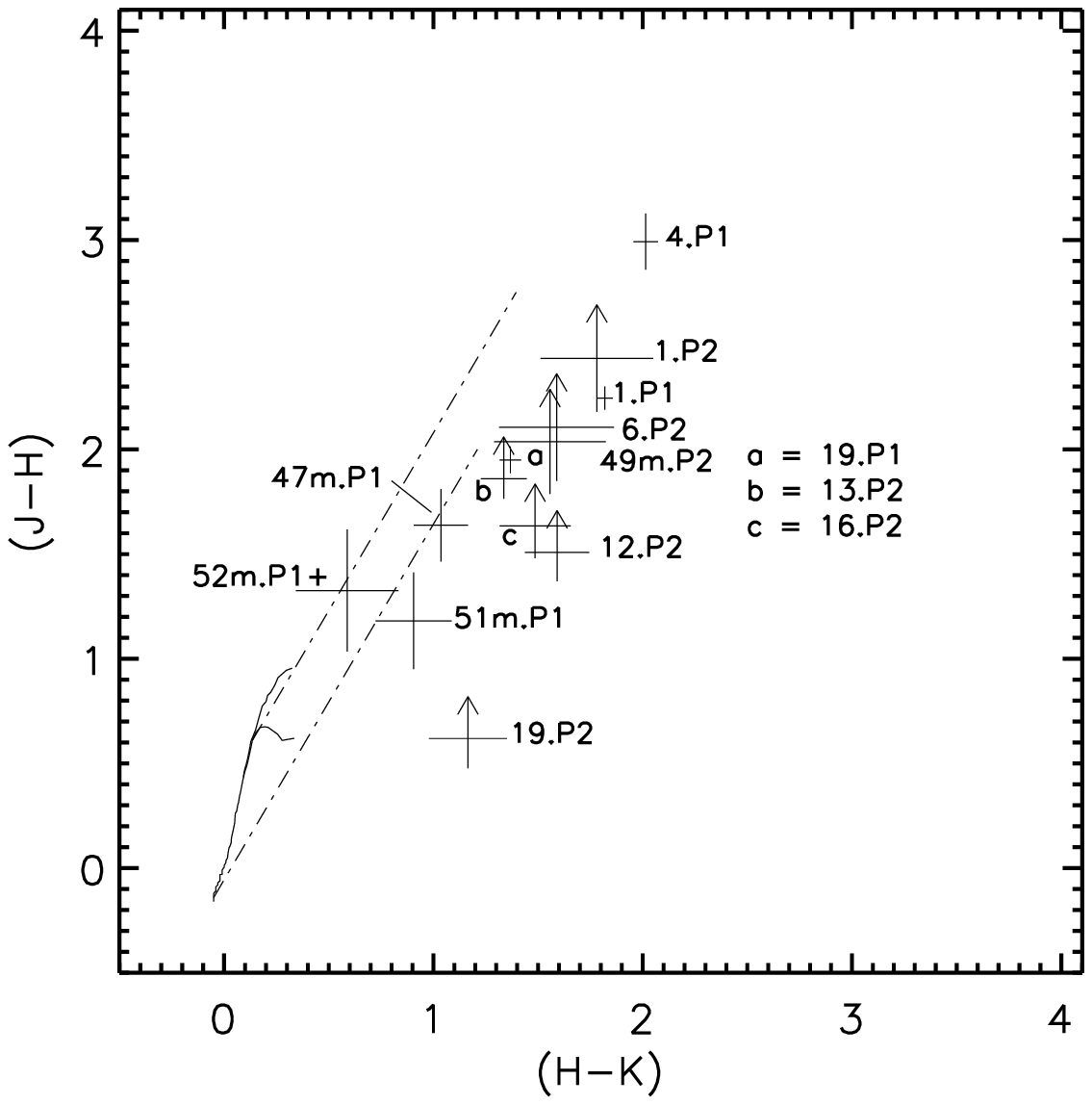}
\caption{As in Fig.~\ref{fig:ccd_p123} but showing the position of YSOs
  spatially associated with submm clumps. The size of the symbols
  indicate the uncertainty in colour. YSOs are identified using the
  nomenclature from Table~\ref{tbl:ysosubmm}, e.g., the YSO from the
  ``P2'' sample associated with clump~13 is labeled ``13.P2''. A lower
  limit to the \emph{(J$-$H)} colour is indicated by an arrow symbol.
\label{fig:ccyso}}
\end{figure}

\begin{figure}
\includegraphics[width=84mm]{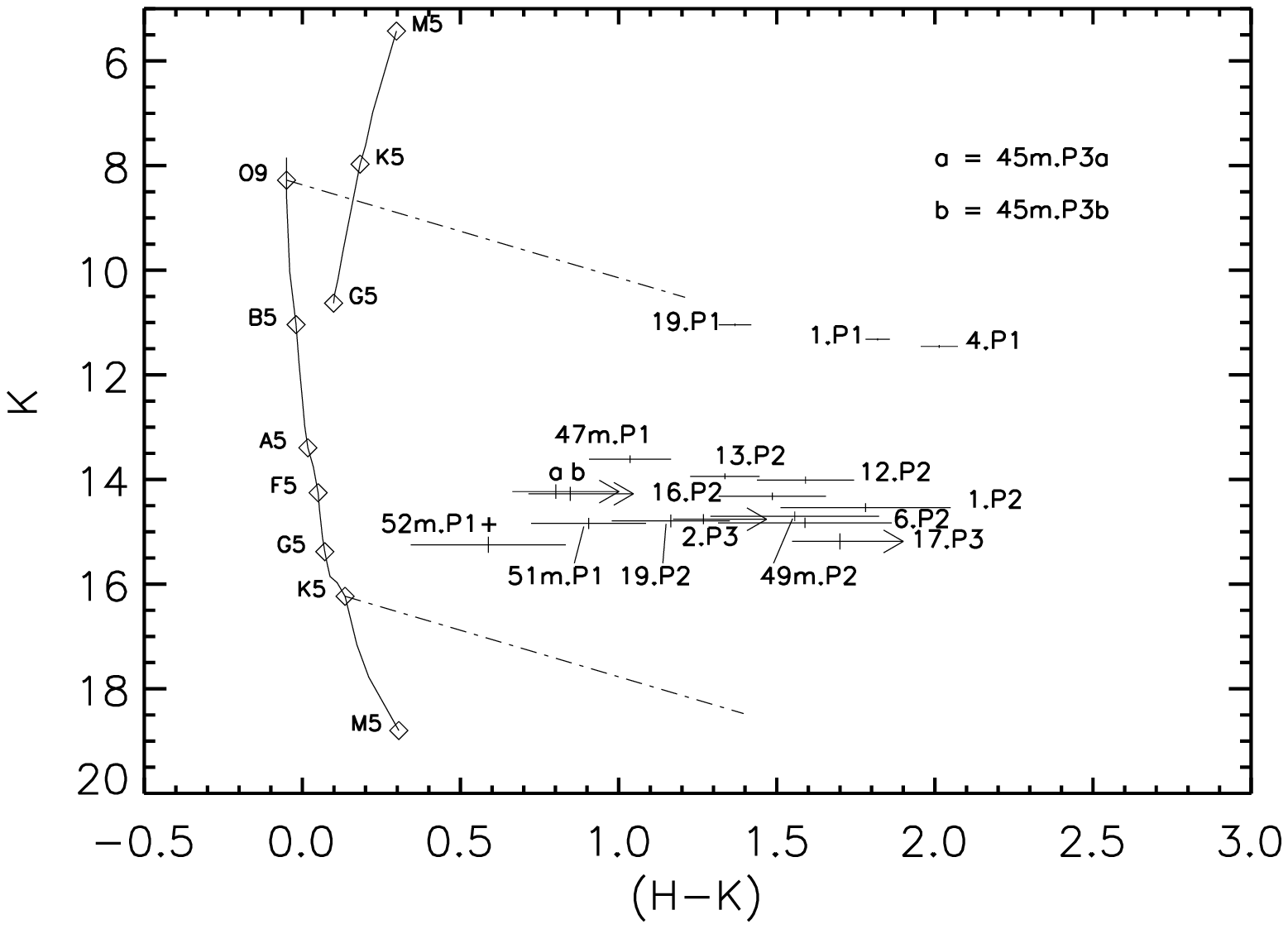}
\caption{As in Fig.~\ref{fig:cmd_p123} but showing the position of YSOs
  spatially associated with submm clumps. The size of the symbols
  indicate the uncertainty in magnitude and colour. YSOs are
  identified using the nomenclature from Table~\ref{tbl:ysosubmm},
  e.g., the YSO from the ``P2'' sample associated with clump~13 is
  labeled ``13.P2''. A lower limit to the \emph{(H$-$K)} colour is
  indicated by an arrow symbol.
\label{fig:cmyso}}
\end{figure}

In Fig.~\ref{fig:ccyso} and Fig.~\ref{fig:cmyso} we plot the CC and CM
diagrams showing the YSOs associated with the submm clumps. The CM
diagram clearly separates the intermediate-mass YSOs (HAeBe stars)
associated with clumps~1, 4 and 19, from the lower luminosity (solar
mass or T~Tauri) YSOs. The former objects provide an important constraint
on the timing of star-formation throughout the molecular cloud as
HAeBe stars clear substantial portions of surrounding molecular material on
timescales of $\sim 10^6$~years \citep{fue98}.

The T~Tauri YSOs found associated with the submm
clumps, and those associated with the dense C$^{18}$O filament (see
\S~\ref{sec:ysodist}), are also likely tracing star formation on similar time
scales. Given that the typical radius for the submm clumps is only
$0.25$ pc and the velocity dispersion of T~Tauri stars is typically
1--2 km~s$^{-1}$ \citep{neu98} it will only take the T~Tauri star
0.2--0.4 Myr to no longer be associated with the clump as seen by an
observer (assuming, for example, that the star is moving at 45$\degr$ to the
line-of-sight). Similarly YSOs associated with the C$^{18}$O filament
have not had enough time to disperse far from their point of origin.

There is growing observational evidence from the analysis of
large-scale maps of molecular clouds that star-formation is more
likely to occur in high column density regions of molecular
clouds where a lower ionization fraction can reduce the efficiency of
magnetic support \citep{mck99} and/or turbulent magnetohydrodynamic
support \citep{ruf98}. Many studies have found an extinction threshold
($A_{V}^{t}$) where regions with $A_V > A_{V}^{t}$ are able to collapse and
fragment to form stellar-sized structures; observed
values range from $A_V \sim 4$ for Taurus \citep{oni98}, through $A_V
\sim 6$ for Perseus \citep{kir06} to values around $A_V \sim 15$ for
Ophiuchus \citep{joh04}. The \citet{hat05} study of
Perseus does not show such a clear threshold, but does find an increasing
probability of finding cores at increasing levels of $A_V$. We note
that all of these studies refer to the conditions required for core
formation and do not consider the YSO content of the core.

\begin{figure}
\includegraphics[width=84mm]{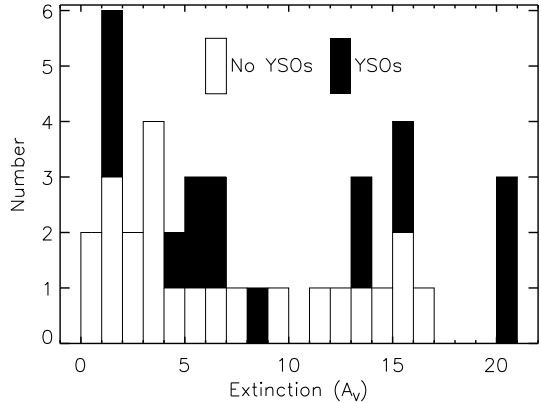}
\caption{Plot showing the number of clumps with (black) and without (white)
  YSOs as a function of clump extinction for the combined \citet{ker01}
  and \citet{moo07} sample. For example, there are 5 clumps with $1
  \leq A_V < 2$ of which 2 have associated YSOs and 3 do not. The rightmost
  bin includes all clumps with $A_V \geq 20$. 
\label{fig:av_combi}}
\end{figure}

In Fig.~\ref{fig:av_combi} a histogram shows how the number of
clumps with and without associated YSOs varies as a function of
$A_V$. While we do not see an obvious single threshold
value it does appear that star-formation is very unlikely to occur in
the very low column density clumps. There are 14 clumps with $A_V <
4$, of which 11 (79\%) do not have associated YSOs and it is unlikely
that star formation will occur in these clumps. At the other end
of the extinction range all three of the clumps with $A_V > 20$ have
YSOs. The remaining 22 clumps with $4 < A_V < 20$ are essentially
evenly split, with 10 having associated YSOs and 12 not having
YSOs. Based on the studies of other star-forming regions mentioned
above we interpret the clumps in this extinction range without YSOs as
being \emph{potential} star-formation sites. The lack of detectable
YSOs with these clumps could reflect their relative youth, or it may
be the case that any YSOs that have formed may be very low mass
and thus fall below the 2MASS detection limits even for our P3
sample.

\begin{table*}
\centering
\begin{minipage}{160mm}
\caption{YSOs associated with submm clumps.}
\label{tbl:ysosubmm}
\begin{tabular}{lccccccc}
\hline
               
               & \multicolumn{4}{c}{2MASS YSO sample} &       &   PDR (P) \\
Clump ID$^{a}$ & P1    & P1+   & P2     & P3          & $A_V$ & or Isolated (I)\\
\hline
1         & 02210449+6106045$^b$       & $\cdots$                &  02210217+6105496      & $\cdots$                & 5.1  & I \\
2         & $\cdots$                   & $\cdots$                & $\cdots$               &  02205994+6107173       & 1.6  & I \\
3$^{c}$   & $\cdots$                   & $\cdots$                & $\cdots$               & $\cdots$                & 4.5  & I \\
4         & 02205329+6108215           & $\cdots$                & $\cdots$               & $\cdots$                & 15.4 & I \\
6         & $\cdots$                   & $\cdots$                &  02203428+6107425      & $\cdots$                & 13.1 & P \\
12        & $\cdots$                   & $\cdots$                &  02195506+6105211      & $\cdots$                & 68.2 & P \\
13        & $\cdots$                   & $\cdots$                &  02195384+6102192      & $\cdots$                & 1.4  & P \\
16        & $\cdots$                   & $\cdots$                &  02194584+6101426      & $\cdots$                & 1.1  & P \\
17        & $\cdots$                   & $\cdots$                & $\cdots$               &  02193257+6107399       & 8.7  & P \\ 
19        & 02192184+6107069           & $\cdots$                &  02192272+6107157      & $\cdots$                & 21.7 & P \\ 
45m       & $\cdots$                   & $\cdots$                & $\cdots$               &  02214126+6105457       & 37.7 & I \\
          &                            &                         &                        &  02214085+6105439       &      &   \\
47m       & 02215369+6106275           & $\cdots$                & $\cdots$               & $\cdots$                & 6.8  & I \\
49m       & $\cdots$                   & $\cdots$                &  02215837+6106405      & $\cdots$                & 5.4  & I \\
51m       & 02220660+6107253           & $\cdots$                & $\cdots$               & $\cdots$                & 6.5  & I \\
52m       & $\cdots$                   & 02222067+61073111       & $\cdots$               & $\cdots$                & 15.7 & I \\
53m$^{d}$ & $\cdots$                   & $\cdots$                & $\cdots$               & $\cdots$                & 13.4 & I \\
\hline
\end{tabular}

$^{a}$ \citet{moo07} ID numbers have a trailing m (e.g. 7m), other ID numbers from \citet{ker01}\\
$^{b}$ 2MASS 02210449+6106045 = MSX6C G133.5492+00.0900 \\
$^{c}$ \emph{IRAS}~02171+6058 \\
$^{d}$ \emph{IRAS}~02186+6033 = MSX6C G133.6890+00.1643 \\
\end{minipage}
\end{table*}

\subsection{General YSO Spatial Distribution} \label{sec:ysodist}

Visually there appear to be four concentrations of YSOs in the KR~140
molecular cloud (for discussion we will denote these Regions~A through
D) along with a noticeable lack of YSOs in the northeast portion of the
molecular cloud (see Fig.~\ref{fig:p123dist}). Region~A is at the
eastern end of the molecular filament near $2^{\rmn{h}} 22^{\rmn{m}}$,
Region~B is associated with submm clump~1 at  $2^{\rmn{h}}
21^{\rmn{m}}$, Region~C is near the central PDR at $2^{\rmn{h}}
20^{\rmn{m}}$, and finally Region~D is associated with clumps~17, 18
and 19 near $2^{\rmn{h}} 19^{\rmn{m}} 30^{\rmn{s}}$. To quantify this
visual impression we divided the molecular cloud area into $2\arcmin
\times 2\arcmin$ boxes and counted the number of stars falling within
each box as a rough measure of the YSO surface density. The average
number of YSOs/box is $0.7\pm1.4$ ($\pm1\sigma$) and boxes near the
four regions in question had surface densities of 7, 7, 6 and 11 for A
to D respectively. There were no other regions in the molecular cloud
with similar $>3\sigma$ deviations from the average value.

Region~A consists of 17 YSOs, four of which are associated with submm
clumps. All of the YSOs in this region are T~Tauri-like with an
average \emph{(H$-$K)} value of $0.92\pm0.25$. None of the YSOs are
coincident with \emph{IRAS}~02186+6033 (MSX6C~G133.6890+00.1643),
strengthening its identification as a Class~0/I YSO.

The focal point of Region~B is the submm clump KMJB~1 which was
identified as a Class~I YSO (\emph{IRAS}~02174+6052) in \citet{ker01}.
The higher resolution of the \emph{MSX} and 2MASS data reveal that 
Clump~1 contains two YSOs, including an intermediate-mass HAeBe
which is also a \emph{MSX} point source with a very red mid-infrared
spectrum (F$_{8.3}=0.22$ Jy, F$_{14.65}=0.73$ Jy). Just
to the east of the clump there are two very red
YSOs. 2MASS~02210851+6105582 is detected only in one 2MASS band with
\emph{K}$=14.306$ and a lower limit colour of \emph{(H$-$K)$_L$}
$>3.26$. It is also a bright \emph{MSX} point source
(MSX6C~G133.5561+00.0919) with an extremely red spectrum (F$_{8.2} =
0.34$ Jy, F$_{14.65} = 1.4$ Jy). The other YSO,
2MASS~02210610+6106043, is detected in both the \emph{H} and \emph{K}
bands and has \emph{(H$-$K)} $=3.63$ and \emph{K} $=11.899$. Both of
these YSOs are probably more highly embedded versions of 1.P1. In
addition, just to the west of KMJB~1, there is another highly embedded
YSO with \emph{K} $=15.208$ and \emph{(H$-$K)$_L$} $>2.30$. It is
quite likely that these objects have formed together as a single
grouping of stars.

Region~C is interesting because of its association with the PDR ridge seen
in the \emph{MSX} images. The majority of the stars are found to the
northeast of the PDR, while 5 of the remaining 6 stars are found along
the PDR with one of the YSOs being coincident with a submm clump. This
pattern suggests that star-formation is progressing from northeast to
southwest with the submm clumps found along the PDR being the latest
round of star formation.

Region~D contains the submm clumps~17, 18, and
19. Intermediate-mass and solar-mass YSOs are associated with clump~19
and, just to the north of this clump, there are an additional six
T~Tauri YSOs. All three of these clumps make up the infrared source
\emph{IRAS}~02157+6053 which \emph{MSX} and 2MASS now clearly show is
actually a concentration of YSOs combined with extended emission from
the PDR.

\section{Discussion} \label{sec:discuss}

\subsection{Spontaneous versus Sequential Star Formation} \label{sec:sssf}

When examining star formation processes within a molecular cloud
containing an \mbox{H\,{\sc ii}} region, a key distinction is
between sequential star formation, where the trigger for star
formation is directly related to the presence of
the \mbox{H\,{\sc ii}} region \cite*[e.g.,][]{deh05}, and spontaneous
star formation, which occurs with no apparent causal relationship with
the \mbox{H\,{\sc ii}} region. It is well known that for both
classical density-bounded \mbox{H\,{\sc ii}} regions and
blister/champagne-flow \mbox{H\,{\sc ii}} regions, the ionization
front (IF) is preceded by a shock front as it propagates into the
molecular cloud \citep{ost85,bed81}. Time-dependent models of PDR/IF
development around OB stars \citep*{rog92,dia98} show that the IF and
PD front merge rapidly, i.e., on time scales less than the
main-sequence lifetime of the exciting star, for typical molecular
cloud densities. With this in mind it is useful to consider two
distinct environments within the molecular cloud; the region beyond
the direct influence of the \mbox{H\,{\sc ii}} region, and the PDR,
which is the region directly influenced by the expansion and UV flux
of the \mbox{H\,{\sc ii}} region and which can be clearly traced by
PAH emission. A comparison of the YSO population found in each of
these regions then gives a measure of the relative importance of
sequential and spontaneous star formation within the molecular cloud.

To delineate these two regions we use the $1.5\times10^{-6}$
W~m$^{-2}$~sr$^{-1}$ contour in the \emph{MSX} Band~A image to
(generously) define the extent of the PDR. We find that 42\% of the
full YSO sample (P1, P1+, P2 and P3) lie within the PDR as defined and
58\% lie outside the PDR. The exact percentages will vary somewhat
depending on how the PDR is defined. For example, using a more
conservative definition for the PDR region ($1.75\times10^{-6}$
W~m$^{-2}$~sr$^{-1}$) results in 31\% of the YSOs within the PDR and
69\% outside the PDR. The percentages, in both cases, are the
same for the YSO population both with and without including the P1+ sample.
In all cases it is clear that a substantial fraction of the YSO
population lies beyond the PDR in this star-forming region and is thus
\emph{not} the result of sequential star formation associated with the
\mbox{H\,{\sc ii}} region.

\subsection{A Scenario for Star Formation in KR~140}

The exciting star of the \mbox{H\,{\sc ii}} region, VES~735, is about
2 Myr old \citep{ker99}, and kinematic models of the \mbox{H\,{\sc
    ii}} region set expansion timescales of 1-2 Myr \citep{bal00}. Our
observations and analysis of KR~140 are consistent with a molecular
cloud undergoing continuous star formation over this period. There is
no clear signature of a significant older population of stars seen in
the 2MASS photometry of the region (see
Figs.~\ref{fig:ccd_p123} and \ref{fig:cmd_p123}).  Lower mass YSOs that
formed at the same time as VES~735 have had time to disperse through
the molecular cloud and may form some of the more spatially
distributed population of YSOs seen in our sample.

The expansion of the  \mbox{H\,{\sc ii}} region has swept up material which has
subsequently collapsed to form the submm clumps that we observe in the
PDR. At the same time the elongated morphology of the densest portion
of the molecular cloud has resulted in ongoing star formation in regions
well away from the direct influence of the \mbox{H\,{\sc ii}}
region. As discussed earlier, the association of HAeBe and T~Tauri
stars with dense regions traced in the submm and C$^{18}$O support
ages of $\la 10^6$ years. The presence of the Class 0/I YSOs,
\emph{IRAS}~02171+6058 and \emph{IRAS}~02186+6033 shows there
continues to be very recent star formation activity on the timescale
of $\sim 10^5$ years \citep{rob06}.

Finally, many of the massive, apparently starless, submm clumps
associated with the PDR may represent the next sites of star formation
in the molecular cloud. Deeper images of the region at longer
wavelengths would be able to determine if these clumps are starless
and test the scenario outlined above by being able to distinguish
between different YSO evolutionary stages \citep{rob06}.

\section{Conclusions} \label{sec:conclude}

While much of the focus of studies of star formation in and around
\mbox{H\,{\sc ii}} regions is on the interface regions, our study of KR~140
illustrates that spontaneous star formation occurring away from the
interface regions in these molecular clouds can be as important as
sequential or triggered star formation related to the expansion of the
\mbox{H\,{\sc ii}} region. Our analysis shows that the molecular cloud
associated with KR~140 contains an extensive population of YSOs that
is not associated with the PDR. 

We have also shown that this YSO population is not uniformly distributed though
the molecular cloud but is concentrated in four distinct regions. Many of the
YSOs are also clearly associated with a filamentary structure within
the molecular cloud which is traced by C$^{18}$O emission. Finally, we
see that both low-mass and intermediate-mass YSOs are forming throughout the
molecular cloud. Particularly interesting is the region around KMJB~1
(\emph{IRAS}~02174+6052) which contains a tight grouping of a number
of embedded intermediate-mass and solar-mass YSOs including the very
highly embedded intermediate-mass YSO 2MASS~02210610+6106043.

\section*{Acknowledgments}
This research has made extensive use of the NASA/IPAC Infrared
Science Archive (IRSA), the Canadian Astronomy Data Centre (CADC), and
data from the Two Micron All Sky Survey (2MASS). IRSA is operated by
the Jet Propulsion Laboratory, California Institute of Technology,
under contract with NASA. 2MASS is a joint project of the
University of Massachusetts and the Infrared Processing and Analysis
Center/California Institute of Technology, funded by NASA and the
NSF. The CADC is operated by the National Research Council of Canada
with the support of the Canadian Space Agency.  FCRAO was supported by
NSF grant AST 0540852. CB is supported by an RCUK Fellowship at the
University of Exeter, UK. The authors would like to acknowledge the work
done by Caroline Pomerleau in the initial stages of this project
supported by the Women in Engineering and Science program of NRC Canada.

\end{document}